\RequirePackage{lineno}
\documentclass[twocolumn,prb]{revtex4-1}
\usepackage{graphicx}
\usepackage{epsfig}
\usepackage{dcolumn}
\usepackage{bm}
\usepackage{amsmath}
\usepackage{amsbsy}
\usepackage{amssymb}
\usepackage[usenames]{color}
\usepackage{ulem}

\newcommand{\ket}[1]{\ensuremath{\,|{#1}\rangle}}

\begin{document}


\setlength\linenumbersep{1.5mm}

\title{Efficient spin control in high-quality-factor planar micro-cavities}

\author{G. F. Quinteiro$^1$, P. Dmitruk$^1$, A. A. Aligia$^2$}
\affiliation{$^1$Departamento de F\'{\i}sica and IFIBA, FCEyN, Universidad de Buenos Aires,
Pabell\'on 1,Ciudad Universitaria, 1428, Buenos Aires, Argentina\\
$^2$Centro At\'{o}mico Bariloche and Instituto Balseiro, Comisi\'{o}n Nacional
de Energ\'{\i}a At\'{o}mica, 8400 Bariloche, Argentina}

\date{\today}

\begin{abstract}
%
A semiconductor microcavity embedding donor impurities and excited by a laser field is modelled. By including general decay and dephasing processes, and in particular cavity photon leakage, detailed simulations show that control over the spin dynamics is significally enhanced in high-quality-factor cavities, in which case picosecond laser pulses may produce spin-flip with high-fidelity final states. 

\end{abstract}

\maketitle


\section{Introduction}

Photons and excitons (X) can be made to strongly interact in high-quality cavities containing a semiconductor quantum well, leading to a repetitive coherent exchange of energy between the two particles.\cite{Wei92} When the energy exchange ocurrs faster than the decay time of the individual components, a combined state, the exciton-polariton, is said to have formed. 

Exciton-polaritons show a variety of features, that motivate studies in multiple directions.
Current interest in exciton-polariton research in 2D micro-cavities mainly focus on its liquid state and non-equilibrium Bose-Einstein condensation (BEC) covering many aspects of the problem\cite{BEC}.
The interaction of polariton fields with impurities or defects has been studied for the case of a polariton fluid scattered by centers acting on the photonic component of the field\cite{Carusotto04}, a polariton gas scattered by spin-independent disorder potential acting on the exciton degree of freedom\cite{Sav97}, and the scattering of polaritons from spinless impurities acting on the excitonic component of the field\cite{Ste86}. 
However, to the best of our knowledge, there are no reported studies on the interaction of a polariton field with a single spin degree of freedom.

\begin{figure}[h]
  \centerline{\includegraphics[scale=.4]{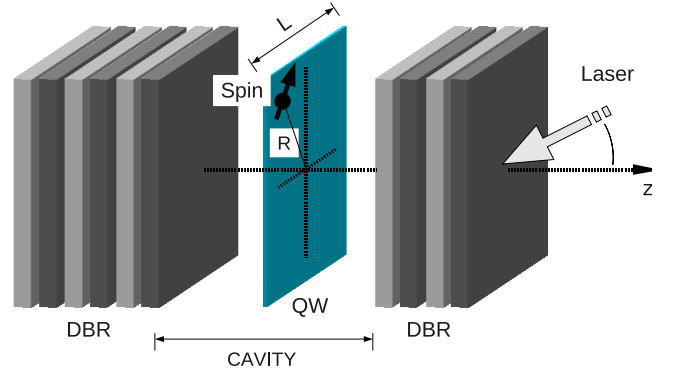}}
  \caption{Pictorial representation of the system. Two Distributed Bragg Mirror (DBR) structures, placed at the sides of a quantum well (QW), confine photons injected from outside by a laser. The photons produce QW excitons that interact with the impurity spin localized at position $\mathbf R$.}
  \label{fig:system}
\end{figure}

Here we study the dynamics, including relaxation processes, of a diluted exciton/photon field interacting with a single impurity of spin $s=1/2$.
The system is depicted in Fig.\ \ref{fig:system}. It consists of a 2D photon cavity embedding a quantum well (QW), which contains few donor impurities \cite{Khitrova-1, Yamamoto-01}. The whole system is assumed to be at low temperature and excited by a laser from outside. 
We show that the quantum control of a single spin is more efficient for high-quality-factor cavities. Thus, a spin-flip in a high-fidelity final state could be produced with a single laser pulse of a few picoseconds. Since typical decoherence times for impurity spins in semiconductors are in the $\mu$s time-scale, the system can act as a high speed quantum memory or qubit\cite{Qui08, Chi05, And11, Krout04}. We believe that the present proposal and that for the implementation of two-qubit polariton-induced operations\cite{Quinteiro-01, Puri12} suggest that a complete quantum-computing scalable architecture based on a solid-state system is possible using polaritons in 2D-microcavities.


\section{Coupling with the environment}

Real systems cannot be entirely isolated from their environment. This is specially true for solid-state systems where several particles co-exist. When control is in mind, undesired interactions make the evolution unpredictable with the possibility of partial or total failure of the control operation.
In our case, excitons, photons, and spins suffer from the coupling with the environment.

For highly pure samples, low temperatures and low exciton density, the relevant decoherence processes for excitons are those causing spin flip of electrons and holes, with conversion between bright and dark excitons \cite{Rui-1, Murayama-1}.  In general hole spins loss coherence faster than electron spins; for instance, in CdTe the spin relaxation of electrons is 29ps\cite{Murayama-1}, while that for holes is $<7$ps\cite{Akas04}.
In addition, the annihilation of excitons must be also considered, with associated lifetime of hundred ps in GaInNAs/GaAs \cite{Lu-1}.

While different processes, such as structural disorder\cite{Savona-1} are responsible for the loss of photon population and coherence, the main process is photon leakage off the cavity due to its finite Q-factor, which leads to a lifetime of the order of $\tau = 15$ps \cite{Cerna-01}.

At extremely low concentration of impurities with densities $n_I\simeq10^{13}$ cm$^{-3}$, electrons bound to different donors are well localized and do not interact among them \cite{Fu-01}. The interaction with the nuclei is dominant (due to the strong confinement of the localized state). At temperatures $T < 10 K$ the transverse relaxation time T2$^*$ is a few ns for the electron bound to a donor, and the spin relaxation time is of the order of $\mu$s for donors in GaAs~\cite{Fu-01}.

\section{Hamiltonian}
\label{Sec:Model}

In what follows, we work in the Heisenberg picture; thus, time-dependent operators shall be everywhere understood. The free Hamiltonian reads
\begin{equation}\label{Eq_H0}
    H_0 = \sum_{\alpha \, \mathbf k}
    \varepsilon_{\mathbf k}\, \hat{b}_{\alpha \mathbf k}^\dag \hat{b}_{\alpha \mathbf k} +
    \sum_{\chi\, \mathbf k}
    \hbar \omega_{\mathbf k}\, \hat{c}_{\chi \mathbf k}^\dag \hat{c}_{\chi \mathbf k}~,
\end{equation}
where the first and second terms correspond to excitons  ($\hat{b}_{\alpha \mathbf k}^\dag/\hat{b}_{\alpha \mathbf k}$) and cavity photons ($\hat{c}_{\chi \mathbf k}^\dag/\hat{c}_{\chi \mathbf k}$).
The QW quantum confinement splits the heavy- and light-hole electronic bands, forming excitons out of conduction-band electrons with total angular momentum $j_z=1/2$ and valence-band heavy holes with $j_z=3/2$. Bright ($j_z=1$) and dark ($j_z=2$) excitons are included: $\alpha=\{1,2,3,4\}=\{\uparrow \Uparrow, \downarrow \Uparrow,  \uparrow \Downarrow, \downarrow \Downarrow\}$, where the single (double) arrow identifies an electron (hole) angular momentum. 
 The respective dispersion relations for excitons and photons are $\varepsilon_{\mathbf k} = \varepsilon_0 + (\hbar \mathbf k)^2/2m^*$ and $ \omega_{\mathbf k} = c/n \, (\mathbf k^2+\mathbf k_z^2)^{1/2}$, where $ \mathbf k$ is the in-plane momentum; the momentum $\mathbf k_z$ in the growth direction  is determined by parameters of the cavity, $n$ is the index of refraction, and $c$ the speed of light. The polarization of the photon is $\chi=\{1,2\}$. The ground state energy of the donor is set to zero.

The system is excited by a classical laser field producing photons that propagate inside the cavity.
Using the quasi-mode approximation (useful for high Q-factor cavities)\cite{quasimode}, the cavity-laser interaction reads
\begin{equation}\label{Eq_HLC}
    H_{LC}=\hbar \sqrt{A} \sum_{\chi \,  \mathbf k} \,
    {\cal{V}}_{\chi \mathbf k} (t)
    \, e^{i (\Omega_{\mathbf k} - \bar \Omega ) t}\, \hat{c}_{\chi \mathbf k} + H.c.~,
\end{equation}
where ${\cal{V}}_{\chi \mathbf q} (t)$ is the coupling constant, $A=L^2$ is the area of the system, $\Omega_{\mathbf k}$ is the laser frequency and $\bar \Omega$ a constant adequately chosen to ease the numerical solution, see below.


Cavity photons interact with excitons according to
\begin{eqnarray}\label{}
    H_{L}
&=& \sum_{       \chi \, \alpha \, \mathbf k     }
    g_{\alpha \chi \mathbf k}(\omega_{\mathbf k}) \,
    \hat{c}_{\chi \mathbf k} \, \hat{b}_{\alpha \mathbf k}^\dag + H.c.\,,
\end{eqnarray}
where $g_{\alpha \chi \mathbf k}(\omega_{\mathbf k}) = 0$ for $\alpha = 1,4$. 

The QW contains donor impuritites. The assumption is made that, at low temperature, each impurity has an electron bound to it that contributes a spin $s=1/2$; in addition, the concentration of donors is low enough to ensure that excitons/polaritons will interact only with one selected impurity, located at position $\mathbf R$, when the laser spot is small enough.
\cite{foot1}
Via Coulomb exchange, the electrons belonging to the exciton and the donor impurity  interact through $ H_{XS} =  H_{XS}^{(+)} + H. c.$
\begin{eqnarray}\label{Eq:HXS}
    H_{XS}^{(+)} \hspace{-1mm}
&=& \hspace{-3mm} 
			\sum_{
          \scriptsize
          \begin{array}{cc}
           \mathbf k \, \mathbf k'  \\
          \end{array}
        }
    \frac{  j_{ \mathbf k \mathbf k'}}{A} \,  
    e^{-i({\bf{k - k'}})\cdot \mathbf R}
 \nonumber \\ 
&&
     \times 
     \, \hat{ \mathbf s} \cdot
			\left[
			\frac{\hbar}{2} \hspace{-2mm}
			\sum_{
          \scriptsize
          \begin{array}{cc}
           \chi \, \chi' \, \eta  \\
          \end{array}
        }
        \hat{b}_{(\chi \eta) \mathbf k}^\dag \,  
        \boldsymbol \sigma_{\chi \chi'} \,     
    \hat{b}_{(\chi' \eta) \mathbf k'} \right]
\end{eqnarray}
where the vector spin operator $\hat{\mathbf s} = (\hat{s}_x, \hat{s}_y, \hat{s}_z)$, $j_{ \mathbf k \mathbf k'} = j_0 [1 + a_I^{*\, 2} (\mathbf k- \mathbf k')^2]^{-1/2}$, and $a_I$ is a measure of the impurity electron localization\cite{Boiko10}. Here we adopted a more detailed notation for the spin $\alpha$ of the exciton: the electron (hole) spin has index $\chi$ ($\eta$), and $\boldsymbol \sigma$ is the vector of Pauli matrices.\cite{Qui08}
X-X interaction is disregarded, because we will study the case of low exciton concentration, where $n_X a_B^{*\,2}/A < 1$, with $a_B^{*}$ the exciton Bohr radius.


\section{Method}
\label{Sec:Method}

Different theoretical tools are employed to solve problems in exciton-polariton research. 
Heisenberg equations of motion describe the dynamics of mean values of either exciton/photon operators or polariton operators \cite{Carusotto04, Cerna-1, Sarchi-1}.
It is also common the use of the Gross-Pitaevskii equation\cite{Ciuti-3, Shelykh-1}.
Other methods have also been used, such as  the Hartree-Fock-Popov\cite{Sarchi-2}.

We make use of the Heisenberg equations of motion (HEM) $ \hbar{d\langle \hat{{\cal O}} \rangle}/{dt} = i   \langle\left[ H,\hat{{\cal O}} \right]\rangle $
for mean values  ($ \langle \ldots \rangle$) of operators describing separately excitons, photons, and the impurity spin. This allows us to treat the cases of weak ---where no polaritons exist--- and strong coupling, as well as to include easily spin-flip processes that cause a polariton to dissociate into a dark exciton and a photon. 

In general, the HEM comprise a set of infinitely coupled equations, that can be ordered in a heirarchy, much as the Bogoliubov-Born-Green-Kirkwood-Yvon (BBGKY) hierarchy of classical statistical mechanics. 
One must then set equations for products or correlation of increasing number of operators. 
In order to close the system of equations, a truncation of the hierarchy is necessary.
We use the truncation scheme $\langle \hat{{\cal O}}_1 \hat{{\cal O}}_2 \rangle = \langle \hat{{\cal O}}_1 \rangle \langle \hat{{\cal O}}_2  \rangle$.
It is important to note that the photon-exciton coupling is not affected by the truncation scheme; therefore, the formation of polaritons (strong-coupling regime) is accurately described.

The system-bath coupling is properly introduced in the HEM by the formalism of Quantum Heisenberg-Langevin equations, which leads to additional terms in the equations: damping, Lamb shift, and stochastic force $\cal F$.
A simpler way to deal with the environment is by introducing constants, taken from experiments or other theoretical works, directly in the HEM, taking into account the results of the detailed microscopic derivation.
Here we follow the phenomenological procedure, by adding constants directly in the HEM \cite{Carusotto04, Savasta96}.  
Photons are coupled to the radiation field at zero temperature outside of the cavity, resulting in the addition of a term $- \xi_q c_{\chi \mathbf q}$; the damping $\xi_q$ becomes very large when $q$ is such that the normal component of the field exceeds the critical angle separating low and high DBR reflectivity.
For excitons we introduced a term $ - \beta_\alpha b_{\alpha \mathbf k}$, with a spin-dependent constant $\beta_\alpha$, accounting for radiative recombination and spin flip (no scattering is considered).
For the impurity spin, a general constant $\gamma$ is used in all component, since no extenal magnetic field exists to distinguish among them. For long times, the spin relaxes, but does not vanish; thus, an equilibrium state is defined.

We consider a circularly-polarized laser field that excites $\hat{b}_{2 0}^\dag$, and an impurity located at $\mathbf R = 0$.
To simplify the calculations, we eliminate fast oscillations by moving to a rotating reference frame, with frequency $\bar \Omega$, setting $\langle \hat{{\cal O}} \rangle =   e^{-i \bar \Omega t} \, \langle \hat{{\cal O}}' \rangle $, for ${\cal O} = c, b$. For the sake of simplicity, we hereafter denote the rotating frame version $ \langle \hat{{\cal O}}' \rangle$, simply as  ${\cal O}$. The equations of motion read
\begin{eqnarray}
\label{Eq:HEM_Sz}
    \frac{ds_{z}}{dt}
&=&
	 \hspace{-2mm}
    -\gamma \bar s_{z}
    +\frac{\hbar}{A}
    \sum_{
          \scriptsize
          \begin{array}{cc}
            \mathbf k \mathbf k'  \\
          \end{array}
    }
    j_{\mathbf k \mathbf k'}^{+} 
   \left(
        s_{y}{\rho}_{1 \mathbf k, 2 \mathbf k'}^{x}
        +s_{x}{\rho}_{1 \mathbf k, 2 \mathbf k'}^{y}
     \right)  \\
\label{Eq:HEM_Sx}
    \frac{ds_{x}}{dt}
&=&
   \hspace{-2mm}
     -\gamma  s_{x}
    -\frac{\hbar}{A}
    \sum_{
          \scriptsize
          \begin{array}{cc}
            \mathbf k \mathbf k'  \\
          \end{array}
    }
     j_{\mathbf k \mathbf k'}^{+}
     \left(
      s_{y}{\rho}_{1 \mathbf k, 1 \mathbf k'}^{z} 
          +s_{z}{\rho}_{1 \mathbf k, 2 \mathbf k'}^{y}
     \right) 
     \\
\label{Eq:HEM_Sy}
    \frac{ds_{y}}{dt}
&=&
    \hspace{-2mm}
    -\gamma  s_{y}
    +\frac{\hbar}{A}
    \sum_{
          \scriptsize
          \begin{array}{cc}
            \mathbf k \mathbf k'  \\
          \end{array}
    }
    j_{\mathbf k \mathbf k'}^{+}
    \left(
         s_{x}{\rho}_{1 \mathbf k, 1 \mathbf k'}^{z}
        -s_{z}{\rho}_{1 \mathbf k, 2 \mathbf k'}^{x} 
    \right)
\end{eqnarray}
where $ \bar s_{z} =  s_{z}-s_{z \infty} $, ${\rho}_{n \mathbf k, n \mathbf k'}^{z} =  ( {\rho}_{n \mathbf k, n \mathbf k'} -  {\rho}_{n+1 \mathbf k, n+1 \mathbf k'})/2$, ${\rho}_{n \mathbf k, m \mathbf k'}^{x} = ({\rho}_{n \mathbf k, m \mathbf k'} +  {\rho}_{m \mathbf k, n \mathbf k'})/2$ and ${\rho}_{n \mathbf k, m \mathbf k'}^{y} = i ( {\rho}_{n \mathbf k, m \mathbf k'} -  {\rho}_{m \mathbf k, n \mathbf k'})/2$,\cite{quin-pierma05} with ${\rho}_{n \mathbf k, m \mathbf k'} = b_{n \mathbf k}^* b_{m \mathbf k'}$ and $j_{\mathbf k \mathbf k'}^{+} = j_{\mathbf k \mathbf k'} + j_{\mathbf k' \mathbf k}$. 
\begin{eqnarray}
\label{Eq:b1q}
    {d {b}_{1 \mathbf  q} \over dt} 
&=&
     - \left[
     		\beta_{1}  
     		+ i  \left( \frac{\varepsilon_{\mathbf q}}{\hbar} - \bar \Omega  \right) 
      \right] {b}_{1 \mathbf  q}
    +\beta_{12} {b}_{2 \mathbf  q}
         \\
&&
\hspace{-5mm}	 
    -  \frac{i}{2 A}\sum_{\mathbf k}
    j_{\mathbf q \mathbf k}^{+}
    \left(
            s_- \, {b}_{2 \mathbf k}
            + s_z  \, {b}_{1 \mathbf k}
    \right) 
    \nonumber     \\
\label{Eq:b2q}
   {d {b}_{2 \mathbf  q} \over dt} 
&=&
     - \left[
     		\beta_{2}  
     		+ i  \left( \frac{\varepsilon_{\mathbf q}}{\hbar} - \bar \Omega  \right) 
      \right] {b}_{2 \mathbf  q}
    + \beta_{12} {b}_{1 \mathbf  q}
\nonumber \\
&&
	 \hspace{-7mm}	 
    -  \frac{i}{\hbar} \sum_\chi g_{2 \chi \mathbf q} \, 
		{c}_{\chi \mathbf q}
    -   \frac{i}{2 A} \sum_{\mathbf k }
    j_{\mathbf q \mathbf k}^{+}
   \left(
           s_+ {b}_{1 \mathbf k}
        - s_z {b}_{2 \mathbf k}
    \right)  \vspace{3mm}
\end{eqnarray}
where $s_{\pm} = s_x  \pm i s_y$. 
Similar equations hold between ${b}_{2 \mathbf  q} \leftrightarrow {b}_{3 \mathbf  q}$
and ${b}_{1 \mathbf  q} \leftrightarrow {b}_{4 \mathbf  q}$.
\begin{eqnarray}
   {d {c}_{\chi \mathbf q} \over dt} 
&=&
    - \left[ \xi_{q}  +  i ( \omega_{\mathbf q} - \bar \Omega ) 
    \right] {c}_{\chi \mathbf q}
    -  \frac{i}{\hbar} \sum_{\sigma } g_{\sigma \chi \mathbf q}^* \,
    		{b}_{\sigma \mathbf q} \nonumber \\
&&
	 \hspace{-5mm}	 
    - i \sum_{\sigma \mathbf k} \sqrt{A} 
		{\cal{V}}_{\sigma \mathbf k}^*(t) e^{- i (\Omega_k - \bar \Omega) t }
     		\delta_{\sigma \chi} \delta_{\mathbf k\,\mathbf q}
\,.
\end{eqnarray}
%

\section{Results}

Numerical solution of the HEM is obtained using a 4th-order Runge-Kutta method in a 2D grid of $N \times N$ modes in momentum space. 
Basic units are $\{$meV, ps, nm$\}$, and we use data compatible with GaAs\cite{Berger}.
The values of the different parameters are taken, in most cases, directly from experimental or theoretical work, only ${\cal{V}}_{0}$ and $j_{0} $ are adjusted using our calculations. 
When $g_{\alpha' \alpha \mathbf q}=0$, $b_{2 \mathbf q}(0) \neq 0$ and $s_z(0)=\hbar$, the system of equations becomes linear, and can be solved exactly. $j_0$ is then adjusted to yield a negative eigenvalue that matches the reported binding energy of excitons to donors (about $1$meV). The value so obtained for $\hbar^2 j_0/A \simeq 10^{-5}$ meV is in agreement with previous reports\cite{Chi05, Quinteiro-01}.
We fix the value of the coupling ${\cal{V}}_{0}$ by demanding that the total exciton density $n_X = \sum_{i \mathbf q} b^\dag_{i \mathbf q}b_{i \mathbf q}$ be low, i. e.:  $r=n_X a_B^{*2}/A<1$, so that the X-X interaction can be neglected.

We studied the evolution of spin components, exciton and photon populations when the system, represented by $N = 50$ modes (larger $N$s do not change the result significantly), is excited by a circularly polarized normal-incidence monochromatic laser-pulse ${\cal V}_{\sigma \mathbf k} = {\cal V}_0 \exp\{-(t-t_p)^2/w^2\}$. Cases with and without decoherence are considered.

It is instructive to analyze first the (idealized) decoherence-free case ---no plot presented. We find neither $b_{3\mathbf q}^*b_{3\mathbf q}$ nor $b_{4\mathbf q}^*b_{4\mathbf q}$ populations, while $s_z$ and $b_{1\mathbf q}^*b_{1\mathbf q}$ change little from their initial values. The small change in $s_z$ can be understood as follows: according to Eqs.\ (\ref{Eq:HEM_Sz}) $ds_z/dt \propto (\hbar j_0/A) b_{1\mathbf q}^* b_{2 \mathbf q}$ and to Eqs.\ (\ref{Eq:b1q}) $d{b_{1 \mathbf q}}/dt \propto (\hbar j_0/A) b_{2 \mathbf q}$, that roughly yields $ds_z/dt \propto (\hbar j_0/A)^2 b_{2 \mathbf q}^* b_{2\mathbf q }$. This, compared to $ds_x/dt \propto (\hbar j_0/A) b_{2  \mathbf q}^* b_{2  \mathbf q}$, is a very small quantity given the choosen value of $\hbar j_0/A \simeq 10^{-5}$ ps$^{-1}$.
On the contrary, the spin projection in the $xy$-plane can rotate several cycles depending on the temporal width and intensity of the pulse. 
As it is well known from quantum optics, once the laser is turned off, there is a remaining oscillating population of excitons and photons. For certain values of the pulse parameters, these populations are so small ($r \rightarrow 0$) that cannot produce important changes in the spin. 

\begin{figure}[h]
  \centerline{\includegraphics[scale=.7]{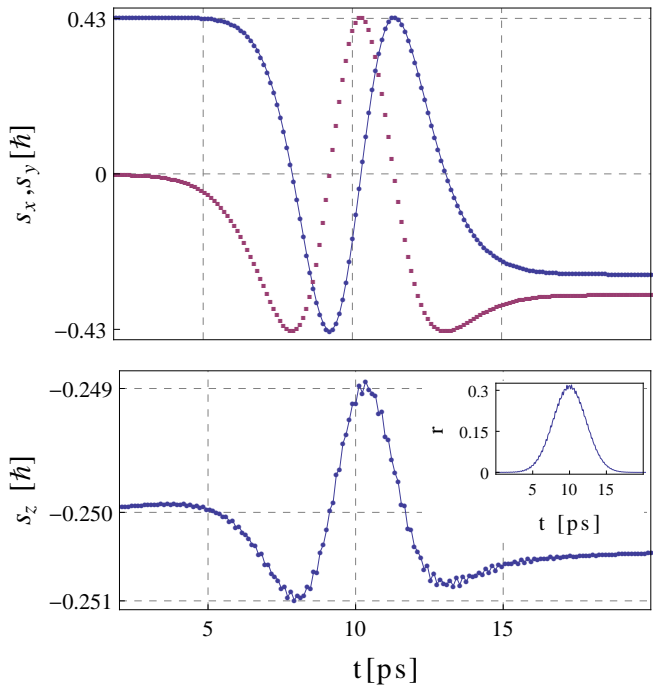}}
  \caption{(color online) Evolution under the excitation by a laser pulse of width $w \simeq 4.5$ps and ${\cal V}_0=25$ps$^{-1}$nm$^{-1}$; the initial state is a spin having mean values $\{s_x=\sqrt{3}\hbar/4, s_y=0,s_z=-\hbar/4\}$.
  Upper Panel: Spin components $s_x$ (solid blue) and $s_y$ (red). 
 Lower Panel: Spin component $s_z$.Inset: fraction $r=n_X a_B^{*2}/A$. }
  \label{fig:S_osc}
\end{figure}

When decoherence is included, there is conversion to dark states $b_{4\mathbf q}$, due to hole spin-flip, and to a lesser extent due to electron spin-flip. Because of the long life-time of these dark states, the fraction $r$ remains finite (though very small compared to its peak value).
Fig.\ \ref{fig:S_osc} presents the results for a simulation with parameters $\{ \Omega_ 0 =2270.$ps$^{-1}, \bar \Omega=2301.2$ps$^{-1}, \varepsilon_0/\hbar=2301.2$ps$^{-1}, \xi_0=6.6\, 10^{-2}$ps$^{-1}, \beta_1=\beta_4=0, \beta_2=\beta_3=10^{-2}$ps$^{-1}, \beta_{12}= \beta_{34}= 3\, 10^{-2}$ps$^{-1}, \beta_{13}= \beta_{24}= 1$ps$^{-1} \}$.
We find that if ${\cal V}_0 < 32$ps$^{-1}$nm$^{-1}$ then $r<1$ and the neglect of the X-X interaction is justified. Under this condition, we see that a single inversion $s_x \rightarrow -s_x$ can be realized in few picoseconds. 
Faster spin motion is observed when the laser intensity (and so the photon/exciton populations) increases. 
As predicted in the previous paragraph, during the whole evolution, the change in $s_z$ is very small, as seen in the lower panel of Fig.\ \ref{fig:S_osc}. In addition the population of dark excitons is also very small compared to that of bright excitons: with the definition $r_i =  b_{i\mathbf q}^*b_{i\mathbf q} a_B^{*2}/A$, we obtain at $t=10$ps  $\{r_1 \simeq 0.3, r_2 \simeq 1.5 \times 10^{-6}\}$ and at $t=20$ps $\{r_1 \simeq 6 \times 10^{-6}, r_2 \simeq 10^{-6}\}$.
%

\subsection{Spin rotation for strong and weak coupling}

The addition of decoherence allows us to address the regimes of strong and weak coupling, and in particular to study the effect that cavity losses have in the spin control.  Weak coupling is characterized by $|\xi_0-\beta_2|>2g/\hbar$ (in our case $2g/\hbar \simeq 2.2$ ps$^{-1}$), and this regime can be simulated by increasing the photon losses of all modes (increasing $\xi_0$), which amounts to considering different cavities with varying quality factor Q.
Two notes of caution: First, we have treated the laser-photon coupling in the quasimode approximation, valid for high-quality-factor cavities. Therefore, we will refrain from studying cases with large values of $\xi_0$.
Second, the laser-photon coupling ${\cal V}_0$ is, in general, affected by changes in the photon losses $\xi_0$; however, we can envisage situations where one can increase $\xi_0$ without affecting ${\cal V}_0$, for example --but not exclusively-- by reducing only the reflectivity of the left DBR in Fig.\ \ref{fig:system}.
%

\begin{figure}[h]
  \centerline{\includegraphics[scale=.95]{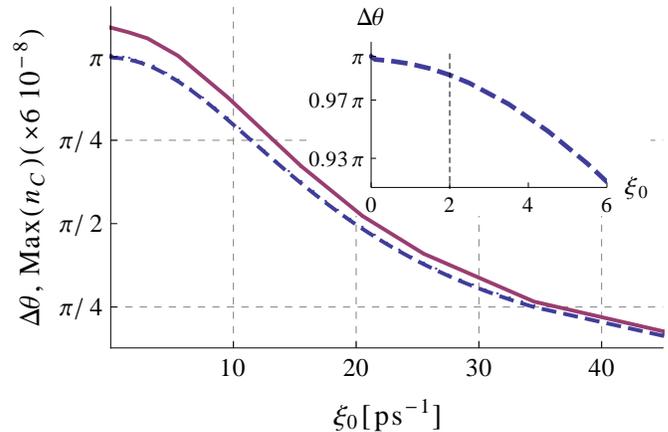}}
  \caption{(color online) Degree of spin in-plane rotation $\Delta\theta$ (dashed blue) as a function of the cavity photon loss, and maximum cavity-photon population (solid red). Inset: Zoom-in of rotation angle $\Delta\theta$ for low $\xi_0$.}
  \label{fig:S_PhEscape}  
\end{figure}

Fig \ref{fig:S_PhEscape} shows the effect that the increase in photon leakage, at fixed laser field intensity, has on the rotation of the impurity spin. For simplicity other sources of decoherence are disregarded. For all simulations, we used one set of laser parameters \{$w\simeq 4.5$ps, ${\cal V}_0=15$ps$^{-1}$nm$^{-1}$\} for a gaussian pulse which produces, {\it without} photon loss, a rotation from the initial state $s_x=\sqrt{3}\hbar/4$ to the final state $s_x=-\sqrt{3}\hbar/4$ at $t=15$ps, i.\ e.\ a change in the angle $\Delta \theta = \pi$. Next we simulated situations of increasing $\xi_0$ and plotted $\Delta \theta (\xi_0)$. In addition, we plotted the maximum photon population acchieved during the pulse. We observe that for $\xi_0<2$ ps$^{-1}$ ($Q > 1500$) there is almost full rotation of $s_x$, and that for lower quality-factor cavities (high $\xi_0$) the spin changes little. The population of cavity photons and excitons follow this tendency. 
%

The effect of decoherence can be characterized with the fidelity $F$. If the final state we wish to obtain is the {\it pure} spin state $-1/2\ket{\uparrow}+\sqrt{3}/2\ket{\downarrow}$, having mean values $\{s_x = -\sqrt{3}\hbar/4, s_y = 0, s_z = \hbar/4\}$, the formula for the fidelty reduces to $F=(-1/2\hbar) (\sqrt{3}s_x-s_z-\hbar)$, see Jozsa\cite{Jozsa}. For the cases $\xi_0=3.5$ps$^{-1}$ and $\xi_0=20.5$ps$^{-1}$ the resulting fidelity is $F=0.9965$ and $F=0.607$, respectively.

We interpret the enhanced rotation in high-Q cavities in the following way. For high Q, as seen in Fig.\  \ref{fig:S_PhEscape}, the photon density is larger, and a repetitive and longer interaction with excitons is possible.  This leads concomitantly to the formation of polaritons, with the excitonic component causing impurity spin rotations. In contrast, for lower values of Q, photons tend to leave the cavity faster, and there is small conversion to excitons. As was mentioned before, it is perhaps easier to envisage a cavity, whose Q factor is lowered by degrading the left DBR in Fig.\ \ref{fig:system}. Then, a naive picture tells us that the laser field produces photons inside the cavity at the same rate in the high and moderate Q cases. In the latter, photons are more prompt to leak out and produce less excitons.

\begin{figure}[h]
  \centerline{\includegraphics[scale=.95]{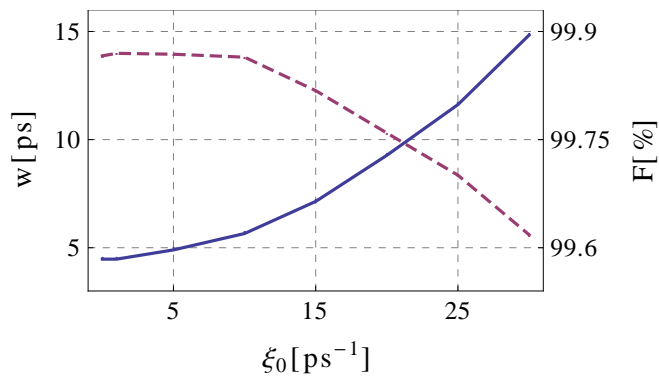}}
  \caption{(color online) Pulse width $w$ (solid blue) required to produce full rotation of the $s_x$ spin component, and corresponding fidelity $F$ (dashed red) as a function of cavity loss $\xi_0$. }
  \label{fig:S_width}
\end{figure}

In addition, we can ask what the pulse width $w$ should be to ensure a full rotation of $s_x$, for different values of cavity loss (see Fig.\ \ref{fig:S_width}). As expected from the previous analysis, we see that one requires longer pulses to produce the rotation, but in contrast to what happened before, the fidelity is almost unchanged. We attribute the behavior of $F$ to the fact that the only source of decoherence is photon loss in these simulations and that the final state is forced (by changing $w$) to be the closest possible to the ideal state.
%

Finally, it is worth mentioning that we have used a conservative value for $j_0$. For example, Puri {\it et al}\cite{Puri12} reports a much higher value of $j_0$ for QDs replacing impurities. This would lead to even faster spin control, together with more efficient control on $s_z$. However, fs laser pulses have a broad frequency spectrum, and may excite several polariton modes. This may lead to destructive interference effects, which may reduce the effectiveness of a large $j_0$.


\section{Conclusions}

We studied the optical control of single spins in micro-cavities accounting for all sources of decoherence. When the system is in the strong-coupling regime, the spin manipulation is most efficient and can be done by a few-picoseconds laser pulse.  
This suggests that single spins embedded in high Q-factor planar cavities can act as  quantum memories and as qubits, with the optical excitation being the mechanism to control the state of the memory or to perform one-qubit operations for quantum computing. This optical control produces high-fidelity final states in very short time: a single operation can be performed $10^6$ faster than the typical decoherence time of the impurity spin qubit (compared to other proposals using for example ion traps\cite{Kirch09}) with a fidelity of $F>99.8\%$.  
We believe that the present proposal for one-qubit operations, together with a previous one for implementing two-qubit operations in the same system\cite{Quinteiro-01, Puri12} show that polaritons in 2D-microcavities is a promising system for the implementation of solid-state quantum-computing scalable architectures.


\section{acknowledgments}

G. F. Quinteiro thanks financial assistance by CONICET and ANPCyT (PICT2007-873), Argentina, and the Fulbright Commission, USA. A. A. Aligia thanks financial assistance by CONICET (PIP No. 11220080101821) and ANPCyT (PICT R1776).



\begin{thebibliography}{}

\bibitem{Wei92}
C. Weisbuch, M. Nishioka, A. Ishikawa, and Y. Arakawa,
Phys. Rev. Lett. {\it 69}, 3314 (1992).

\bibitem{BEC}
F. Tassone, C. Piermarocchi, V. Savona, and A. Quattropani, P. Schwendimann
Phys. Rev. B {\bf 56} 7554 (1997).
%
L. Ferrier, E. Wertz, R. Johne, D. D. Solnyshkov, P. Senellart, I. Sagnes, A. Lemaitre, G. Malpuech, J. Bloch, 
Phys. Rev. Lett. {\bf 106}, 126401 (2011)
%
Iacopo Carusotto, and Cristiano Ciuti,
Phys. Rev. B {\bf 72} 125335 (2005).
%
Iacopo Carusotto, Michiel Wouters, Cristiano Ciuti
J Low Temp Phys (2007) {\it 148}: 459–464.
%
C. Ciuti, P. Schwendimann, B. Deveaud, and A. Quattropani,
Phys. Rev. B {\bf 62}, R4825 (2000).
%
Markus Muller and Joel Bleuse, Regis Andre
Phys. Rev. B {\bf 62}, 16 886 (2000).
%
S. Portolan, O. Di Stefano, S. Savasta, F. Rossi, and R. Girlanda
Phys. Rev. B {\bf 77}, 035433 (2008).
%
S. Savasta, G. Martino, R. Girlanda,
Solid State Comm. {\it 111}, 495 (1999)
%
V. Savona, Z. Hradil, A. Quattropani, and P. Schwendimann,
\textit{Phys. Rev. B} {\bf 49}, 8774 (1994).

\bibitem{Carusotto04} 
Iacopo Carusotto and Cristiano Ciuti,
Phys. Rev. Lett. {\it 93}, 166401 (2004).

\bibitem{Sav97}
V. Savona, C. Piermarocchi, and A. Quattropani, F. Tassone, P. Schwendimann
Phys. Rev. Lett.  {\bf 78}, 4470 (1997).

\bibitem{Ste86}
T. Steiner, M. L. W. Thewalt, E. S. Koteles and J. P. Salerno,
\textit{Phys. Rev. B} \textbf{34}, 1006 (1986).
B. Lavigne and R. T. Cox,
\textit{Phys. Rev. B} \textbf{43}, 12374 (1991).

\bibitem{Khitrova-1}
G. Khitrova, H. M. Gibbs, F. Jahnke, M. Kira, and S. W. Koch,
``Nonlinear optics of normal-mode-coupling semiconductor
microcavities'', Rev. Mod. Phys. {\it 71}(5), 1591-1639 (1999).

\bibitem{Yamamoto-01}
Y. Yamamoto, F. Tassone, H. Cao, \textit{Semiconductor Cavity Quantum
Electrodynamics}, Springer Tracks in Modern Physics {\bf 169}, 2000.

\bibitem{Chi05} 
G. Chiappe, J. Fernandez-Rossier, E. Louis, and E. V. Anda,
Phys. Rev. B {\bf 72}, 245311 (2005)

\bibitem{Qui08}
G.\ F.\ Quinteiro,
Phys. Rev. B {\bf 77}, 075301 (2008)

\bibitem{And11}
J.\ A.\ Andrade, A.\ A.\ Aligia and G.\ F.\ Quinteiro,
J. Phys.: Condens. Matter {\bf 23}, 215304  (2011).

\bibitem{Krout04}
Miro Kroutvar, Yann Ducommun, Dominik Heiss, Max Bichler, Dieter Schuh, Gerhard Abstreiter \& Jonathan J. Finley,
Nature {\bf 432}, 81-84 (2004) 

\bibitem{Quinteiro-01}
G. F. Quinteiro, J. Fernandez-Rossier, and C. Piermarocchi, \textit{Phys. Rev. Lett.} \textbf{97}, 097401 (2006).

\bibitem{Puri12}
Shruti Puri, Na Young Kim, Yoshihisa Yamamoto,
Phys. Rev. B {\bf 85}, 241403(R) (2012).

\bibitem{Ramvall97}
P. Ramvall, N. Carlsson, P. Omling, L. Samuelson, W. Seifert, and Q. Wang
Appl.\ Phys.\ Lett. {\bf 70} 243 (1997).

\bibitem{Oh-1}
I. -K. Oh, Jai Singh, Journal of Luminescence \textbf{85} (2000) 233-246.

\bibitem{Tsi2005}
E. Tsitsishvili, R. v. Baltz, H. Kalt
Phys. Rev B {\bf 72}, 155333 (2005).

\bibitem{Zutic04}
Igor Zutic, Jaroslav Fabian, S. Das Sarma
Rev. Mod. Phys. {\bf 76} 323 (2004).

\bibitem{Murayama-1}
A. Murayama, K. Seo, K. Nishibayashi, I. Souma, and Y. Oka, Applied
physics letters, \textbf{88}, 261105 (2006).

\bibitem{Lu-1}
S. L. Lu, L. F. Bian, M. Uesugi, H. Nosho, A. Tackeuchi, and Z. C.
Niu, Applied physics letters, \textbf{92}, 051908 (2008).

\bibitem{quasimode}
C. W. Gardiner and M. J. Collett, \textit{Phys. Rev. A} {\bf 31},
3761 (1985).
%
B. J. Dalton, Stephen M. Barnett and P. L. Knight, \textit{Journal
Modern Physics} {\bf 46}, 1315-1341 (1999).
%
Hiroshi Ajiki, \textit{J. Opt. B: Quantum Semiclass. Opt.} {\bf 7}
(2005) 29-34.


\bibitem{Akas04}
S.\ Akasaka, et al, Applied Physics Letters {\bf 85} 2083 (2004).

\bibitem{Rui-1}
Rui Shen, Hirofumi Mino, Tasuya Kimukawa, Shojiro Takeyama, Grzegrz
Karczewski, Tomasz Wojtowicz, Jacek Kossut, Phisica E \textbf{22} (2004)
611-614.

\bibitem{Savona-1}
Vincenzo Savona, J. Phys.: Condens. matter \textbf{19} (2007)
295208.

\bibitem{Cerna-01}
R. Cerna, D. Sarchi, T. K. Paraiso, G. Nardin, Y. Leger, M. Richard,
B. Pietka, O. El Daif, F. Morier-Genoud, V. Savona, M. T.
Portella-Oberli, B. Deveaud-Pledran, 'Optical manipulation of the
wave function of quasiparticles in a solid' (no reference
available).

\bibitem{Ciuti-3} Iacopo Carusotto and Cristiano Ciuti,
arxiv:cond-mat/0504554v1 (2005).

\bibitem{Shelykh-1} 
I. A. Shelykh, Y. G. Rubo, G. Malpuech, D. D.
Solnyshkov, and A. Kavokin, \textit{Phys. Rev. Lett.} \textbf{97}
066402 (2006). N. A. Gippius, I. A. Shelykh, D. D. Solnyshkov, S. S.
Gavrilov, Yuri. G. Rubo, A. V. Kavokin, S. G. Tikhodeev, and G.
Malpuech, \textit{Phys. Rev. Lett.} \textbf{98} 236401 (2007). T. C.
H. Liew and I. A. Shelykh, \textit{Phys. Rev. B} \textbf{80}
161303(R) (2009).

\bibitem{Fu-01}
Kai-Mei C. Fu, C. Santori, C. Stanley, M. C. Holland, and
Y. Yamamoto, Phys. Rev. Lett. \textbf{95} 187405 (2005).

\bibitem{foot1}
When quantum computing is in mind, two-qubit operations are performed by illuminating simultaneously two impurities.

\bibitem{Boiko10}
I.\ I.\ Boiko,
Semiconductor Physics, Quantum Electronics \& Optoelectronics, 2010. V. \textbf{13}, N 2. P. 214-220.

\bibitem{Cerna-1} 
R. Cerna, D. Sarchi, T. K. Para\"{\i}so, G.
Nardin, Y. Leger, M. Richard, B. Pietka, O. El Daif, F.
Morier-Genoud, V. Savona, M. T. Portella-Oberli, B.
Deveaud-Pl\'edran, \textit{Phys. Rev. B} \textbf{80} 121309(R)
(2009).
%

\bibitem{Sarchi-1} 
D. Sarchi. and V. Savona, \textit{Phys. Rev. B}
\textbf{77}, 045304 (2008).

\bibitem{Sarchi-2} 
D. Sarchi. and V. Savona, arXiv:0706.3719v1 [cond-mat.mes-hall].

\bibitem{Savasta96}
S. Savasta and R. Girlanda, 
Phys. Rev. Lett. \textbf{77}, 4736 (1996).

\bibitem{quin-pierma05}
G.\ F.\ Quinteiro, C.\ Piermarocchi,
Phys. Rev. B {\bf 72}, 045334 (2005)

\bibitem{Berger} 
Lev. I. Berger,
{\it Semiconductor Materials} (CRC Press, Inc. 1997).


\bibitem{Jozsa} 
Richard Jozsa,
Journal of Modern Optics \textbf{41}, 2315-2323 (1994).

\bibitem{Kirch09}
G.\ Kirchmair, J. Benhelm, F. Zahringer, R. Gerritsma, C. F. Roos, and R. Blatt, Phys. Rev. A \textbf{79} 020304 (2009).
J. Benhelm, G. Kirchmair, C. F. Roos, and R. Blatt, Nature Physics \textbf{4}, 463 (2008)


\end{thebibliography}
\end{document}